\documentclass[aps,prb,twocolumn,letterpaper,showpacs,10pt,footinbib]{revtex4-1}
\usepackage[utf8]{inputenc}
\usepackage{graphicx,color,amsmath}
\usepackage{amssymb}
\usepackage{amsthm}
\usepackage[bookmarks]{hyperref}
\usepackage{umoline}

\newcommand{\be}{\begin{equation}}
\newcommand{\ee}{\end{equation}}
\definecolor{mygreen}{rgb}{0,0.5,0}
\definecolor{myblue}{rgb}{0,0,0.75}
\definecolor{mymagenta}{cmyk}{0,1,0,0.12}

\begin{document}
\title{Non-equilibrium $8\pi$ Josephson Effect in Atomic Kitaev Wires}
\author{C. Laflamme, J. C. Budich, P. Zoller, M. Dalmonte}
\affiliation{Institute for Quantum Optics and Quantum Information of the Austrian Academy of Sciences, 6020 Innsbruck, Austria, and \\ Institute for Theoretical Physics, University of Innsbruck, 6020 Innsbruck, Austria}
\date{\today}
\begin{abstract}
We theoretically study a Kitaev wire interrupted by an extra site which gives rise to super exchange coupling between two Majorana bound states. 
We show that this system hosts a tunable, non-equlibrium Josephson effect with a characteristic $8\pi$ periodicity of the Josephson current. We elucidate the physical mechanism deriving a minimal model for the junction and confirm its quantitative accuracy by comparison to the numerical solution of the full model. The visibility of the $8\pi$ periodicity of the Josephson current is then studied using time-dependent simulations including the effects of dephasing and particle losses. Our findings provide a novel signature of Majorana quasi-particles which is qualitatively different form the behavior of a conventional superconductor, and can be experimentally verified in cold atom systems using alkaline-earth-like atoms.
\end{abstract}


\maketitle

\section{Introduction}
The search for observable signatures that identify exotic states of quantum matter and their fractionalized excitations has become a main focus of research in quantum physics. A paradigmatic example is the hunt for Majorana quasi-particles (MQPs) which exist at the ends of topological superconductors \cite{Kitae01}. First experimental evidence \cite{Mourik2012,Deng2012,Rokhinson:2012aa,Das:2012aa,Williams:2012aa,Nadj-Perge:2014aa} consistent with the presence of MQPs has recently been reported in various superconducting hybrid systems  \cite{Lutchyn:2010aa,Oreg:2010aa,Choy:2011aa}. While the ultimate goal is to probe the existence of non-Abelian anyons such as MQPs by performing controlled braiding operations, several possible fingerprints have been proposed that may be easier to access experimentally.  

A prominent example hallmarking MQPs is the fractionalization of the Josephson effect, which can exhibit a $4\pi$ (half frequency) period due to a non-equilibrium population of excited states that is protected by fermion parity conservation~\cite{Kitae01,Rokhinson:2012aa}. 
However, a similar, though non-protected, fractionalization is also known to occur in conventional s-wave superconductors, due to the presence of accidental mid-gap states \cite{Kwon:2004aa,Sau:2012jk}. In contrast, here we show how a dissipationless, non-equilibrium $8\pi$-periodic Josephson effect occurs when two MQPs are subject to a super-exchange coupling via a controllable energy level interrupting a Kitaev chain, an effect that is not found in s-wave superconductors. In addition, we show how our model can be realised in a system of cold atoms in optical lattices, where isolation from the environment creates an ideal platform for the study of such non-equilibrium phenomena.

\begin{figure}[t!]
\centering
\includegraphics[width=0.95\columnwidth]{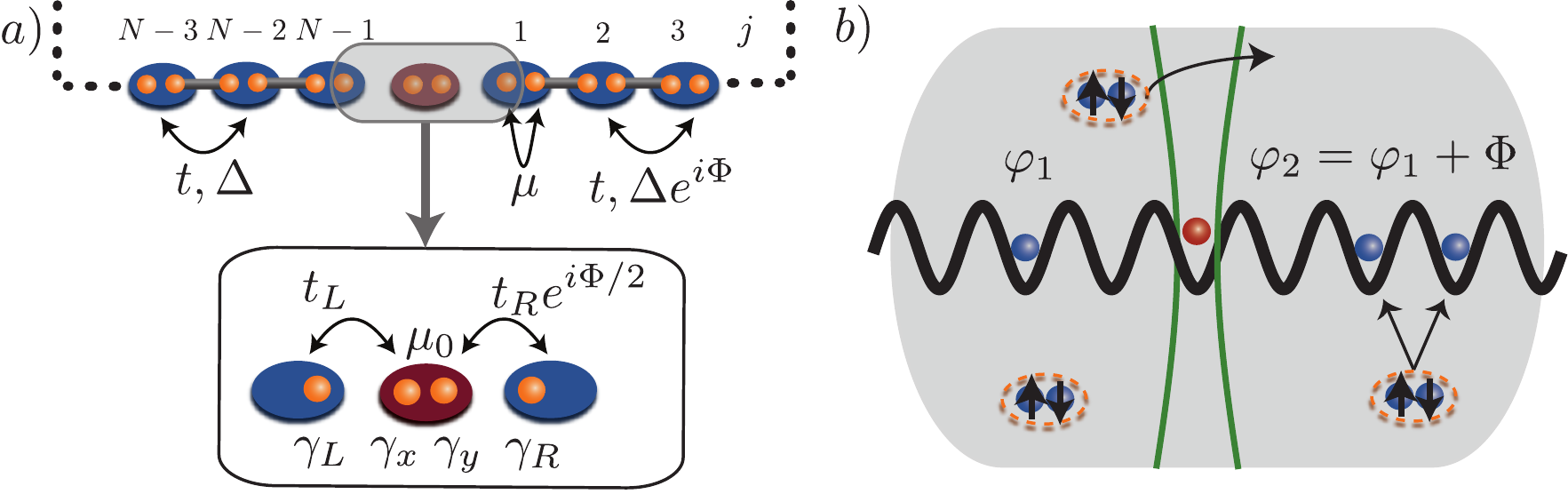}
\caption{\label{fig:one} System Hamiltonian and cold atom setting. a) Schematics of the model Hamiltonian Eq.~\eqref{eqn:tb}: the central part of the system is magnified in the box at the bottom, where the Majorana degrees of freedom included in the simplified model (Eq.~\eqref{eqn:sm}) are highlighted. b) Implementation in a cold atom system. A 1D optical lattice is coupled to a BEC reservoir which gives rise to the Kitaev Hamilonian in the chain. An optical barrier acts both to create the impurity site (red) and triggers the Josephson effect in the reservoir itself. The phase difference across the barrier in the reservoir then acts as the phase $\Phi$ for the optical lattice.}
\end{figure}

Our proposal is motivated by remarkable recent experimental progress with cold atom systems, including the the observation of the non-equilibrium Josephson effect~\cite{Paterno82}, initially demonstrated with Bose-Einstein condensates~\cite{Cataliotti03082001,Levy:2007aa}, and later observed over the BEC-BCS crossover~\cite{Husmann2015,Valtolina2015}. These results demonstrate not only the ability to measure non-equilibrium signals, but in addition, this realisation of the $2\pi$ Josephson effect \cite{Valtolina2015} will provide a crucial piece of our implementation. More concretely, in our proposal, the starting point is an atomic realisation of the Kitaev wire \cite{Jiang:2011aa,Diehl:2011aa, Nascimbne:2013aa,Kraus:2013aa}, here using a system of Alkaline Earth Atoms (AEAs) coupled to a BEC reservoir (see Fig.~\ref{fig:one} b)). AEAs allow the creation of a controllable extra site by means of species dependent potentials \cite{Daley:2008aa}, while the reservoir allows both the implementation of the Kitaev wire, and the modification of the Josephson phase via an underlying Josephson effect of the reservoir itself. In addition, we investigate the visibility of this effect by studying the transient dynamics of the Josephson current in the presence of imperfections, including various dissipation mechanisms (single particle losses and dephasing) captured by a quantum master equation. Our simulations support not only the observability of the $8\pi$ effect, but further underline how this signature is characteristic of Majorana quasi-particles: While
$4\pi$ peaks in the Fourier signal cannot be distinguished from those arising from mid-gap states in an ordinary S-wave SC, and peaks at $4\pi$, $2\pi$ and zero-frequency can be enhanced from dissipation, the $8\pi$ signal visible in our setup provides a signature that cannot be confused with
these undesired effects.

\section{Model and Results}
\subsection{Model Hamiltonian}
We consider spinless fermions with field operators $\psi_j$, where $j=0,\ldots N-1$ labels the sites of a one-dimensional (1D) lattice in ring geometry. The model Hamiltonian reads as
\begin{align}
&H(\Phi)=\sum_{j=1}^{N-1}\left[-t \psi_j^\dag\psi_{j+1}+\Delta \psi_j\psi_{j+1}-\frac{\mu}{2}(\psi_j^\dag\psi_j-\frac{1}{2})\right]\nonumber\\
&+t_L\psi_{N-1}^\dag\psi_0+t_R\psi_0^\dag\psi_1\text{e}^{i\Phi/2}+\frac{\mu_0}{2}\psi_0^\dag \psi_0+\text{h.c.},
\label{eqn:tb}
\end{align}
which describes a proximity induced $p$-wave superconductor \cite{Kitae01} with pairing $\Delta$, interrupted by an extra site at $j=0$ which is assumed to be not affected by the pairing (see Fig. \ref{fig:one} a)). The hopping strength is denoted by $t$ and the chemical potential relative to half-filling by $\mu$. The site at $j=0$ is connected to its neighbors by the hoppings $t_L$ and $t_R$, respectively, and has an energy offset $\mu_0$. The phase factor $\text{e}^{i\Phi/2}$ on the hopping between $j=0$ and $j=1$ models a flux that advances the phase of a Cooper pair by $\Phi$ when moving around the ring.

\begin{figure}[t!]
\centering
\includegraphics[width=0.85\columnwidth]{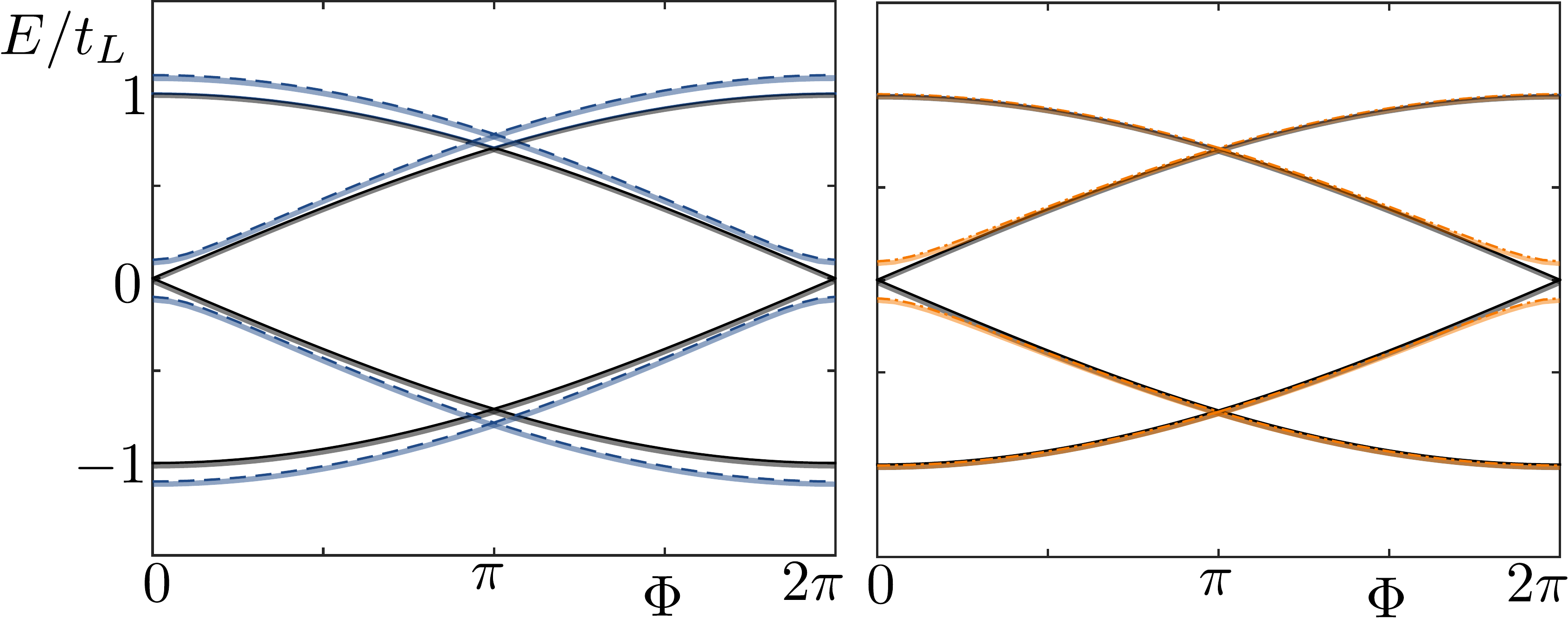}
\caption{\label{fig:two} Energy spectrum of the minimal model in Eq.~\eqref{eqn:sm} with different parameters, with the width of the lines indicating the deviation from the energy spectrum of the full microscopic model (shifted by an energy constant to lie at the same scale) in Eq.~\eqref{eqn:tb} with ($N=10, \Delta = t=10 t_L, \mu=0$). Left Panel: $t_R=t_L, \mu_0=0$ (black, solid), $t_R=1.2 t_L, \mu_0=0$ (blue, dash). Right Panel: $t_R=t_L, \mu_0=0$ (black, solid), $t_R= t_L, \mu_0=0.2 t_L$ (orange, dot-dash). The degeneracies at $\Phi=\pi$ are protected by the global $\mathbb{Z}_2$ parity symmetry, while the degeneracies at $\Phi=0$ are present for $t_L-t_R=\mu_0=0$.}
\end{figure}

For $\lvert\mu\rvert<2t, \lvert\Delta\rvert>0$, and $t_L=t_R=0$,  in the limit of large $N$ the system hosts a single pair of zero-energy MQPs \cite{Kitae01}, $\gamma_L$ and $\gamma_R$, which are localized exponentially around $j=N-1$ and $j=1$, respectively. All other quasiparticles of the superconductor are gapped, such that $\psi_0$ along with $\gamma_L$ and $\gamma_R$ form a subspace that is energetically detached from the bulk spectrum. To understand the qualitative $\Phi$-dependence of Eq. (\ref{eqn:tb}) in the physically relevant regime $t_L,t_R\ll \Delta,t$, we hence consider a minimal model encompassing the dynamics within this low-energy sector. Decomposing $\psi_0$ into the Majorana operators $\gamma_x=\psi_0+\psi_0^\dag,~ \gamma_y=\frac{\psi_0-\psi_0^\dag}{i}$, and setting $\mu_0=0$, the effective Hamiltonian then reads as
\begin{align}
H_J(\Phi)=\frac{1}{2i}\left[t_L\gamma_L\gamma_x-t_R\gamma_R\left(\gamma_x \sin(\Phi/2)+\gamma_y\cos(\Phi/2)\right)
\right].\label{eqn:sm}
\end{align}

In Fig. \ref{fig:two}, we compare the energy spectra of $H_J(\Phi)$ and $H(\Phi)$. The full qualitative agreement confirms that the effective Hamiltonian $H_J(\Phi)$ captures the basic Josephson physics of the full model $H(\Phi)$. To understand the various level (avoided) crossings in Fig. \ref{fig:two}, we first focus on the symmetric case $t_L=t_R$. At $\Phi=0$, we have $H_J(0)=\frac{t_L}{2i}(\gamma_L\gamma_x-\gamma_R\gamma_y)$, i.e., the four Majorana operators form two disjoint pairs giving rise to two single particle (hole) excitations with energy $\frac{t_L}{2}$ ($-\frac{t_L}{2}$). 
The four possible many-body states then have the energies $(-t_L,0,0,t_L)$ which explains the twofold degeneracy at $E=0$. At $\Phi=\pi$, we have $H_J(\pi)=\frac{t_L}{2i}(\gamma_L\gamma_x-\gamma_R\gamma_x)$, i.e., $\gamma_L$ and $\gamma_R$ are coupled to the same Majorana operator $\gamma_x$. 
This gives rise to a zero mode in the single-particle spectrum and the many-body energies are $(-t_L/\sqrt{2},-t_L/\sqrt{2},t_L/\sqrt{2},t_L/\sqrt{2})$ as reflected in the crossings at $\Phi=\pi$ in Fig. \ref{fig:two}. 
At $\Phi=2\pi$, we have $H_J(2\pi)=\frac{t_L}{2i}(\gamma_L\gamma_x+\gamma_R\gamma_y)$, i.e., the analogous situation to $\Phi=0$ but with a sign change of a single-particle excitation energy, reflecting the change of the fermion parity in the ground state \cite{Kitae01}. 
At $\Phi=3\pi$, the situation is analogous to $\Phi=\pi$ with $\gamma_R\rightarrow -\gamma_R$. As for $\Phi=4\pi$, we note $H_J(4\pi)=H_J(0)$.
 However, despite the $4\pi$-periodicity of $H_J$, adiabatically following the ground state in Fig. \ref{fig:two} through the various crossings leads to an $8\pi$-periodic pattern. This is a phenomenon of spectral flow, where the system is pumped to an excited state during one $4\pi$-cycle of the Hamiltonian, and only returns to the initial state after a second cycle.
 
We emphasize that the level crossings in Fig. \ref{fig:two} are of quite different physical nature. The crossings between states with different fermion parity at odd multiples of $\pi$ are robust as long as the fermion parity is conserved. By contrast, the crossings at even multiples of $\pi$ require left/right symmetry and a mid-gap state on the additional site: this is realized by tuning the junction parameters, namely $\mu_0=0$ and symmetric tunnelling $t_L=t_R$. However, tuning of the bulk parameters within the TSC phase supporting the MQPs $\gamma_L,\gamma_R$ is not required as long as the bulk gap is much larger than $t_L, t_R$. In a solid-state setting the decoherence due to the coupling to phonons implies that observing the non-equilibrium population of the unprotected excited state presents a serious challenge. In contrast, in the cold atom setting proposed here, such decoherence channels are not present, thus stabilising these effects. 

Below we describe how the model given in Eq.~\eqref{eqn:tb} can be realised in systems of AEAs trapped in optical lattices, before discussing in more detail the visibility of the $8 \pi$ Josephson effect in the presence of various imperfections. 

\begin{figure}[t!]
\includegraphics[width = 0.95\columnwidth]{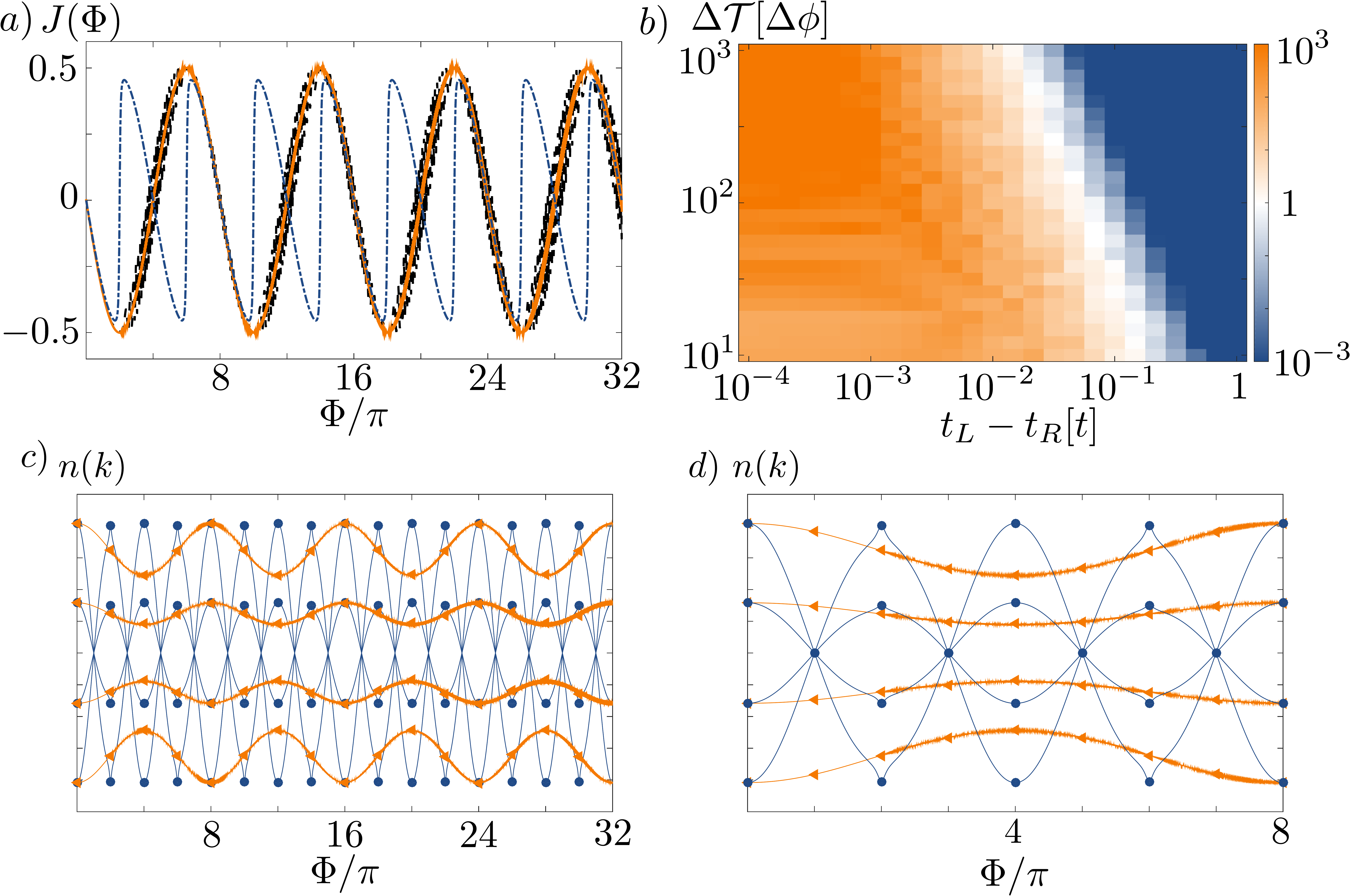}
\caption{ a) Current as a function of $\Phi$ for a system with $N=10$, with parameters $\Delta \mathcal{T}= 10^3\Delta \Phi, t_L-t_R=10^{-4}t$ (orange,solid), $\Delta \mathcal{T}= 10 \Delta \Phi, t_L-t_R=10^{-2}t$ (black, dot-dash) and $\Delta \mathcal{T}= 10^3\Delta \Phi, t_L-t_R=10^{-1}t $(blue, dash).
b) Logarithm of the ratio of the height of the $8\pi$ and $4\pi$ peak of the FFT of the current profile over a range of model parameters. c)  Time-evolution of the some of the $k$-components of the momentum distribution as a function of $\Phi$ for a system with $N=10$, with parameters $t=10t_L, \Delta = 10.1t_L, \Delta \mathcal{T}=10^3 \Delta \Phi$ and $t_r-t_l=10^{-4} t_L$ (orange / triangles) and $t_r-t_l=10^{-1} t_L$ (blue / circles). d): same as in panel c), with the time window $\Phi[0,8\pi]$ magnified.}
\label{fig:CoherentGrid} 
\end{figure} 

\subsection{Experimental realization} 
There are three points required for the realisation of our setup: {\it i)} the implementation of a 1D Kiteav chain, {\it ii)} the addition of the single site separating the two ends of the wire, and {\it iii)} the time-control of the phase $\Phi$. 
In order to address these points in a concrete setup, we consider a system of fermionic AEAs \cite{Schreck09,Killian10,Takahashi07,Schreck11,Takahashi11,Boyd,Foelling14,Ye11,Pagano:2014kq,Cazalilla:2014ty}, trapped in their $^1S_0$ ground state in a 1D lattice. The choice of AEAs allows us to independently trap the $^1S_0$ ground state $|g\rangle$ and the $^3P_0$ metastable excited state atoms $|e\rangle$, which, as we will see, will be of crucial use. 

We first address {\it i)}. While the hopping terms ($t$) arise naturally in the lattice, pairing terms ($\Delta$) can be induced by coupling the fermions in the lattice to a BEC reservoir, where an RF field is used to break up cooper pairs directly into neighbouring sites in the lattice, as described in Ref.~\cite{Jiang:2011aa}.

{\it ii)} We now describe how we can interrupt the chain with a single site. First, at the position $j=0$ a barrier is engineered to inhibit $|g\rangle$ atoms from being at this site, which splits the Kitaev wire into two. This can be done using a highly focused beam at the so called anti-magic wavelength, which acts as a sink for $|e\rangle$, and oppositely on $|g\rangle$~\cite{Daley:2008aa}, resulting in the $|e\rangle$ atoms {\it only} being trapped at this site. Thus the $|e\rangle$ atom at site $j=0$ acts as the additional site coupling the two ends of the wire. While natural hopping into and out of this site is deterred from this barrier, the tunnelling ($t_L$ and $t_R$) are then reintroduced with Raman processes involving a clock transition \cite{Jaksch:2003aa,Gerbier:2010aa,Wall:2015aa}. 

{\it iii)} In fact, the barrier which inhibits $|g\rangle$ atoms to be trapped at $j=0$ also acts as the mechanism which controls the phase $\Phi$. This can be seen as follows. The barrier is turned on via a laser which is highly localised at the $j=0$ position in the optical lattice, but homogenous in the remaining directions and impacts the BEC reservoir, bisecting it into two regions. For a barrier that is only a few times larger than the coherence length of the system, it will act as a thin tunneling barrier between the two regions. 
If the two regions have a different Cooper pair density, an ordinary AC Josephson effect will occur, giving rise to a relative phase $\Phi$ across the junction which oscillates in time \cite{Valtolina2015}. The Josephson frequency $\omega_J$ of this oscillation is proportional to the population imbalance, which constitutes the analog of a bias voltage in the solid state context. Due to the proximity effect, this time dependent phase is inherited by the 1D lattice system, giving rise to the model described in Eq.~\eqref{eqn:tb}. Here $\omega_J$ is on the order of the bare trap frequency and can be controlled via the barrier and reservoir parameters.

Within this setup, there are two main ways to demonstrate the 8$\pi$ periodicity of the Josephson effect by current measurements. First, it is possible to use local interferometric probes, as realised, e.g., in Ref.~\cite{Atala:2014kq}, or to infer the current behaviour from density measurements~\cite{Husmann2015,Valtolina2015}. Second, one can observe clear signatures of the 8$\pi$ periodicity by using the relation between the time-dependent momentum distribution and the current operator~\cite{Mancini25092015,Stuhl25092015}. For the model defined in Eq.~\eqref{eqn:tb}, the relevant current at the junction is defined by:
\begin{equation}\label{cond1}
J (\tau) = \langle i  (\psi_{L-1}^\dagger(\tau) \psi_{0}(\tau) - \text{h.c.})\rangle,
\end{equation}
where $\tau$ denotes the real time on which the Hamiltonian is dependent via the modulation of the phase $\Phi(\tau) = \omega_J \tau$, with the Josephson frequency $\omega_J$, such that  $\Phi(0)=0$.
Since the system we investigate does not display translational invariance, the global current operators cannot be described solely in terms of momentum distribution (momentum is not a good quantum number). Indeed, the total current reads:
\begin{eqnarray}
J &=& (t-t_L-t_R)\sum_{k} \frac{\langle a^\dagger_ka_{k}\rangle\sin(k)}{2} + \nonumber\\
&+& \sum_{k\neq q} \left[(-t_L\langle a^\dagger_ka_{q}\rangle e^{-iq} + \textrm{h.c.}) +\right. \nonumber\\
&+& \left.  (-t_R\langle a^\dagger_ka_{q}\rangle e^{i(k-2q)} + \textrm{h.c.})    \right]
\end{eqnarray}
where the presence of the last two terms reflects the fact that momentum is not a conserved quantity. While these terms are not directly accessible in cold atom experiments, it is possible to identify signatures of the $8\pi$ periodicity via the first term. 

In Fig.~\ref{fig:CoherentGrid}, we show the time-dependent behaviour of the current (a) and the various components of the momentum distribution (c-d) as a function of time in different parameter regimes, for a system of N=10 sites. For system parameters where the current has a dominant $8\pi$ periodicity (the orange line in Fig.~\ref{fig:CoherentGrid} a.), the momentum components $n(k)$ individually mirror this. This is shown in the orange(triangle) line of Fig.~\ref{fig:CoherentGrid} where the identical parameters have been taken. However, when system parameters are such that the current has a dominant $4\pi$ periodicity (the blue dashed line in Fig.~\ref{fig:CoherentGrid} a.), the momentum components reflect this. This is shown with the blue (square) lines of Fig.~\ref{fig:CoherentGrid}c, again with the identical parameters.

We now address the question of the integrity of this $8 \pi$ Josephson effect in our proposed setup subject to imperfections. First, we address the influence of Hamiltonian imperfections $t_L\ne t_R$ as well as $\mu_0 \ne 0$ leading to avoided crossings in the level spectrum at integer multiples of $2\pi$ (see Fig. \ref{fig:two}). We find that Landau-Zener processes restore the $8 \pi$ periodicity of the current at finite bias voltage. Thereafter, we investigate the effect of single particle losses, induced three-body collisions between particles in the wire and pairs in the reservoir, and dephasing in the framework of a Markovian quantum master equation \cite{San-Jose:2012aa,Virtanen:2013aa}.

\section{Time-dependent dynamics}
\subsection{Transport dynamics and $8\pi$ Josephson effect}
We study the current through the junction region at site $j=0$, as defined in Eq.~\eqref{cond1}. In the limit of perfect adiabatic evolution $\omega_J \rightarrow 0$, at the symmetric parameter point $t_L=t_R, \mu_0=0$, the current will be $8\pi$ periodic, as indicated in the dispersion relation (see Fig. {\ref{fig:two}}); any deviation from this fine-tuned parameter point will cause a gap to open and the adiabatic current will be $4\pi$ periodic. However, the $8\pi$-effect is restored at finite $\omega_J$ due to the Landau-Zener effect. This tradeoff between finite $\omega_J$ and finite imperfections is analysed within the coherent time evolution governed by Eq. (\ref{eqn:tb}) in Fig.~\ref{fig:CoherentGrid}, where we numerically calculate the current $J(\tau)$ as a function of time (see Fig. \ref{fig:CoherentGrid}(a)). For small $\omega_J$ and weak imperfections (black solid line), the current displays a clear $8\pi$ periodicity, while increasing imperfections at fixed $\omega_J$ is detrimental (red dot-dashed line). However, larger $\omega_J$ allows the system to follow the avoided crossings due to Landau-Zener tunnelling, thus restoring the $8\pi$ periodicity (blue dashed line).

In order to provide a quantitative picture of the interplay between imperfections and $\omega_J$, we extract the height of the $8\pi$-peak and the $4\pi$-peak from the Fourier transform of the current over a total phase change of $\Phi_T=8\pi$. The ratio of these two peaks is shown in Fig.~\ref{fig:CoherentGrid} panel b) as a function of $\omega_J$ and $(t_L-t_R)$. At intermediate $\omega_J$, the $8\pi$ peak dominates over a wide range of parameters: remarkably, even for imperfections of a few percent, the $8\pi$ signal is still an order of magnitude stronger than that at $4\pi$. This behaviour has been verified with $\Phi _T=32\pi $. Data shown in Fig.~\ref{fig:CoherentGrid} are for $\kappa =0$ with $\Phi _T=8\pi $ to minimise the compound effect of several Landu-Zener crossings (a finite $\kappa $ stabilises this effect and data at $\Phi _T=32\pi $ is shown in these cases, as discussed in the next section).

\begin{figure}[t!]
\includegraphics[width = 0.95\columnwidth]{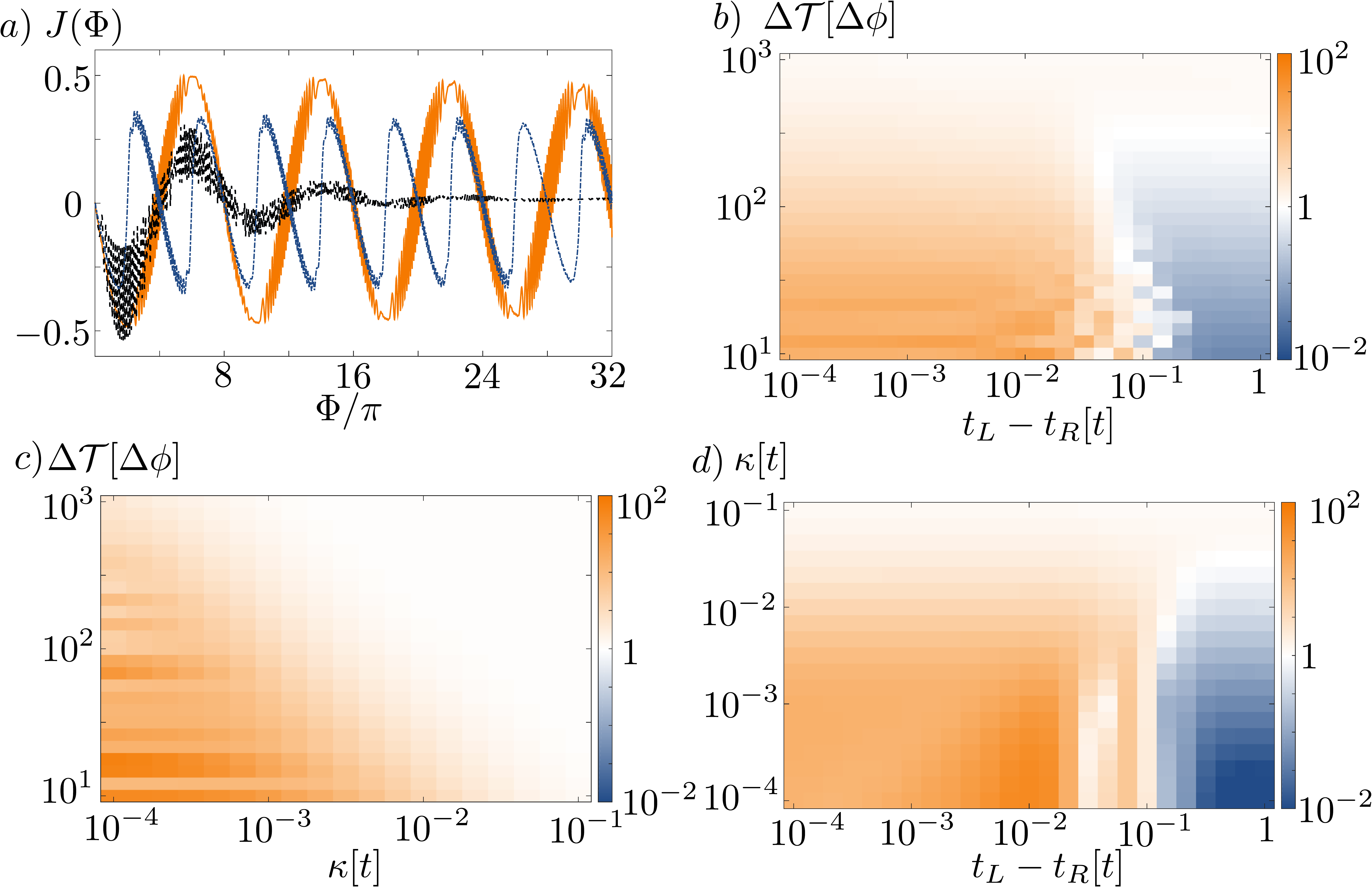}
\caption{a) Current as a function of $\Phi$ for a system with $N=10$, with parameters $\Delta \mathcal{T}= 10\Delta \Phi$  and $\kappa = 10^{-4}t, t_L-t_R=10^{-2}t$ (orange, solid), $\kappa = 5\cdot10^{-3}t, t_L-t_R=10^{-2}t$ (black, dot-dash), $\kappa = 10^{-4}t, t_L-t_R=t$ (blue, dash). b)-d) Ratio of the strength of the $8\pi$ peak of the FFT of the current profile and the $4\pi$ peak with $\kappa = 10^{-3}$ fixed (panel b)), $t_L-t_R = 10^{-2}$ fixed (panel c)) and $\Delta \mathcal{T} = 10 \Delta \Phi$ fixed (panel d)). }
\label{fig:8piGrid} 
\end{figure} 

\subsection{Dissipation and open system dynamics}
In addition to imperfections which cause the system to move away from the symmetric point $t_L=t_R, \mu_0=0$, an experimentally relevant imperfection is due to the coupling of the system to its environment. To account for this, we consider two dissipative channels. The first is a single particle loss at site $j$ with rate $\kappa_j$: in cold atom settings, this represents losses due to inelastic collisions with the background BEC reservoir. The second source of dissipation is dephasing due an effective measurement at rate $\gamma_j$ of the local occupation number $n_j=\psi_j^\dag \psi_j$ by the environment. This is typically represents the effect of spontaneous emission in optical lattice settings. Assuming a weak coupling to a Markovian quantum bath, the time evolution of the system is then governed by the master equation 
\begin{eqnarray}
\partial_\tau \rho &=& 
-\frac{i}{\hbar} \left[ H, \rho \right] 
+ \sum_{j=0}^{L-1}\left[ \kappa_j \mathcal D^{\psi_j}[\rho] +\gamma_j \mathcal D^{n_j}[\rho] \right],
\label{eqn:Lindblad}
\end{eqnarray}
where $\rho$ is the density matrix of the system and the superoperator $\mathcal D^O[\rho]=O \rho O^\dag - \frac{1}{2} \left\{
O^\dag O,\rho \right\}$ is the Lindblad dissipator for an arbitrary Lindblad jump operator $O$. As long as $\gamma_j=0$, Eq. (\ref{eqn:Lindblad}) is still quadratic in the field operators $\psi_j$ and can be solved numerically efficiently. By contrast, $\gamma_j$ leads to quartic terms in the master equation (\ref{eqn:Lindblad}) which we treat in an exact diagonalisation analysis. In what follows, we present results for the full master equation in systems of $N=10$ sites. 

To study the impact of a finite $\kappa_j \equiv \kappa$ on the integrity of the $8\pi$-effect, we numerically solve the master equation~(\ref{eqn:Lindblad}) and calculate the current $J(\tau)$ in the presence of finite loss. In such open settings, the system dynamics is now determined by the competition of three energy scales, corresponding to $\omega_J$, the energy scale related to Hamiltonian imperfections, and $\kappa$. At fixed $\kappa$, one expects a stronger $8\pi$ signal for intermediate $\omega_J$, since  both Landau-Zener tunnelling works at its best even in the presence of imperfections, and dissipation becomes detrimental only after many oscillations periods.

A few examples of the current evolution as a function of time are depicted in Fig.~\ref{fig:8piGrid}a): the main effect of dissipation is to damp the current signal in the system, thus inhibiting transport. However, even for relatively large decay rates (black line, corresponding to decay collision rates of order $\kappa\simeq 1$ Hz~\cite{Jiang:2011aa} to be compared with $t_L\simeq 200$ Hz), the signal stays $8\pi$ periodic for intermediate timescales (combined with a exponentially decaying envelop).  

Following the above analysis, again we quantify the $8\pi$ effect by extracting the ratio of the $8\pi$ and the $4\pi$ peaks from the Fourier spectrum. This ratio is shown for various system parameters and loss rates in Fig.~\ref{fig:8piGrid}b)-d), and illustrates the regimes in which the $8\pi$ signal can be seen. In panel b), we plot the ratio at fixed $\kappa$: the best attainable regime, is for intermediate values of the velocity, where imperfections are relatively harmless up to values on the order of a few percent. In panel c), $t_L-t_R$ is fixed: here, again intermediate speeds work at best, and values of the dissipation of the order of $10^{-2}$ can be tolerated. Finally, in panel d), the speed of the ramp is fixed: the signal is solid in the regime of low losses, and, for intermediate values of imperfections, larger values of the losses, $\kappa$, can be tolerated. The strong signal at these intermediate values of $t_L-t_R$ is consistent with what is expected from Landau-Zener theory, which predicts an optimal tunnelling rate at intermediate gap values in case of finite dissipation and finite speed. 

We have repeated these calculations in the presence of a finite dephasing rate $\gamma$. In this case, the system dynamics is not quadratic in the fermions, so our study was limited to system sizes up to $L=10$ sites. A sample of the results is presented in Fig.~\ref{fig:fig5} a).  Overall, we found that it has qualitatively the same effect as $\kappa$, which can be understood in terms of the protection of the non-equilibrium excited states. While the decay channel $\kappa$ mixes states with different parity the decay channel $\gamma$ mixes states within the same parity, both contributing equally through the evolution from $0$ to $8\pi$. Finally, we have checked how the main effects discussed here are affected by finite-size effects. In the regimes of interest, those effects are negligible at $N=10$. For the $\gamma =0$ case, we have checked this explicitly for some sample points up to $N=30$, while for the $\gamma \not =0$ case, we have systematically checked consistency with the $N=8$ case.

In summary, the $8\pi$ periodicity of the current profile is robust to both the Hamiltonian imperfections and the dissipation considered here. Monitoring the evolution for shorter time (e.g., for a single $8\pi$ cycle) can also substantially improve the signal, as in that case the role of particle losses is less detrimental. 

\subsection{Many body effects} 
\begin{figure}[t!]
\includegraphics[width = 0.95\columnwidth]{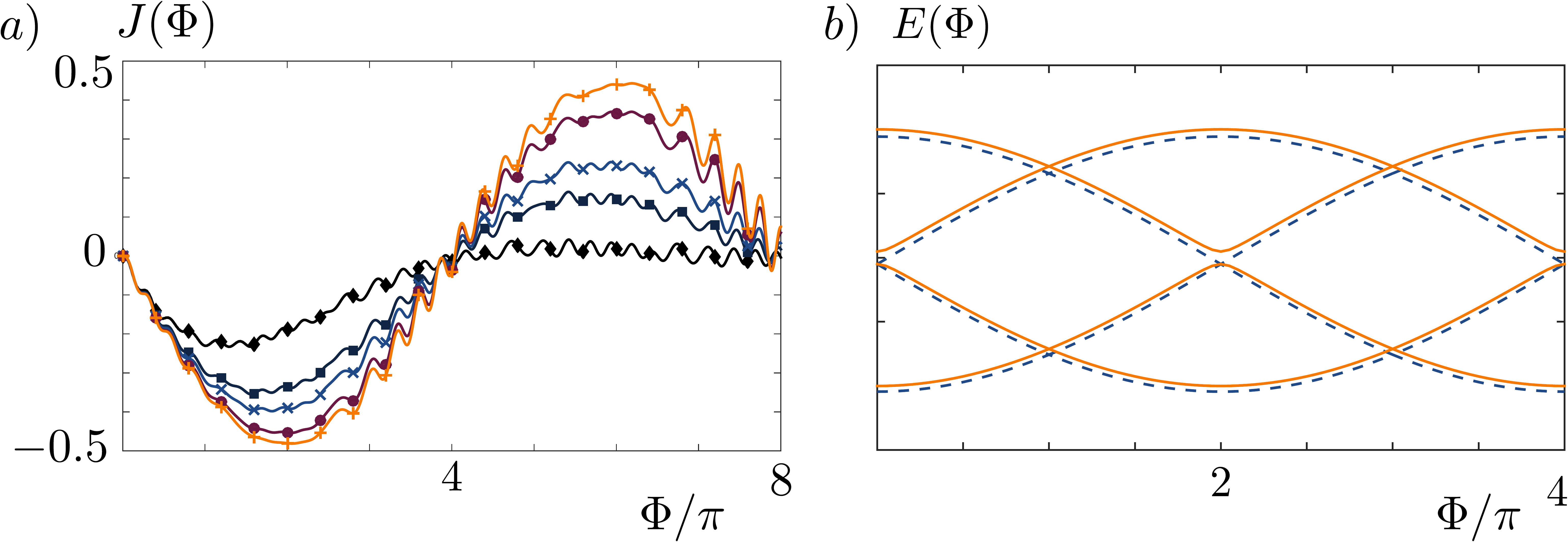}
\caption{a) Current as a function of $\Phi$ for a system with $N=10$, with parameters $\Delta \mathcal{T}= 40\Delta \Phi, t_L-t_R=5*10^{-3}t, \Delta = 1.01t, t_L=0.1t$, and different values of $\kappa=\gamma=[0.02, 0.008, 0.005, 0.002, 0.0008]$ (diamonds, squares, crosses, circles, pluses; ordered top to bottom for short times). b) Energy of the microscopic model in the presence of a nearest neighbour interaction (Eq.~\ref{eq:interaction}). Parameters $\Delta=t=5 t_L, t_R=t_L, \mu=\mu_0=0$, and comparing the non-interacting case (blue/dashed) with a finite interaction $U=0.1t_L$.}
\label{fig:fig5} 
\end{figure}

Finally, we consider the effect of a finite interaction on the energy spectrum of the model. We consider a nearest neighbour interaction of the form
\be
\label{eq:interaction}
H_{\rm int} =  U n_{N-1}n_0+U n_{0}n_1,
\ee
where $n_j=a^\dag_ja_j$ and we assume these are the dominating terms while the interactions are suppressed in the superconducting region. Typically, in systems of AEAs the ratio $U/t\sim 10^{-3}$. As seen in Fig.~\ref{fig:fig5} b) a finite $U$ opens a gap at the level crossings at even multiples of $\pi$. For moderate values of $U$, this effect is then analogous to $t_L\ne t_R$ or $\mu_0\ne 0$.

\section{Conclusions and outlook}
The periodicity of the Josephson effect is closely related to the charge of the particles involved in the tunnelling processes. Intuitively, an $8\pi$-periodicity then corresponds to a fractional charge of $\frac{e}{2}$, which is the physical picture behind the time-reversal protected fractional Majorana fermions discussed in Ref.~\cite{Zhang:2014aa}. In contrast, our model does not involve fractional charges, and our effective Hamiltonian $H_J(\Phi)$ (see Eq.~\eqref{eqn:tb}) is hence $4\pi$-periodic in $\Phi$, in agreement with the Byers Yang theorem \cite{Byers:1961aa}.  The $8\pi$-Josephson effect in our setup is a phenomenon of spectral flow: the system is pumped to an excited state after slowly increasing $\Phi$ by $4\pi$, and returns to the ground state after a second $4\pi$ cycle. Our work thus shows that an $8\pi$-periodic signal can also emerge due to non-protected crossings, analogue to what has been shown to occur for the $4\pi$ effect. However, in the latter case, the accidental $4\pi$ periodicity occurs when the underlying system is a conventional superconductor; here this $8\pi$ effect arrises when the underlying system hosts 'normal' ($\mathbb{Z}_2$) Majorana fermions

We note that while a $12\pi$-periodic Josephson effect has been put forward in the context of two connected quantum wires \cite{Nogueira:2012aa}, we emphasise that these effects are dissipationfull, as there is no controllable gap separating the crossing branches of the Josephson junction from the bulk states.

 {\textbf{Acknowledgements}} We acknowledge useful discussions with M. Baranov, S. Nascimb{\`e}ne, P. Recher, and B. Trauzettel. This work was supported by the ERC Synergy Grant UQUAM, SIQS, SFB FoQus (4016-N23),  and the Austrian Ministry of Science BMWF as part of the UniInfrastrukturprogramm of the Focal Point Scientific Computing at the University of Innsbruck. C.L. is partially supported by NSERC. 

\bibliographystyle{apsrev}

\end{document}